\documentclass[twocolumn,showpacs,preprintnumbers,amsmath,amssymb,superscriptaddress]{revtex4-2}
\usepackage{graphicx}
\usepackage{dcolumn}
\usepackage{bm}
\usepackage{CJK, CJKnumb}
\begin{document}
%\begin{CJK*}{GBK}{song}

\title{Spin-valley polarized edge states and quantum anomalous Hall states controlled by side potential in 2D honeycomb lattices}
\author{{ Wei-Tao Lu$^{1}$, Qing-Feng Sun$^{2,3,4}$, Yun-Fang Li$^{5}$, and Hong-Yu Tian$^{1}$ }\\
\normalsize{${^1}^{}$\emph{School of Physics and Electronic Engineering, Linyi University, 276005 Linyi, China}} \\
\normalsize{${^2}^{}$\emph{International Center for Quantum Materials, School of Physics, Peking University, Beijing, 100871, China}} \\
\normalsize{${^3}^{}$\emph{Collaborative Innovation Center of Quantum Matter, Beijing 100871, China }} \\
\normalsize{${^4}^{}$\emph{Beijing Academy of Quantum Information Sciences, West Bld. $\#$3, No. 10 Xibeiwang East Road, Haidian District, Beijing 100193, China }} \\
\normalsize{${^5}^{}$\emph{School of Mechanical $\&$ Vehicle Engineering, Linyi University, 276005 Linyi, China}} \\}

\begin{abstract}
Based on the tight-binding formalism, we study the effect of side potential on the spin and valley related electronic property of $2$D honeycomb lattices with intrinsic spin-orbit coupling, such as silicene and germanene. The side potential is composed of potential field and exchange field applied on the boundaries of the zigzag nanoribbon. It is found that the side potential could greatly affect the helical edge states with different spin indices and the spin and valley are locked to each other. By adjusting the side potential and ribbon width, the system shows quantum spin-valley Hall effect, valley polarized quantum spin Hall effect, and spin polarized quantum anomalous Hall effect. Due to the side potential and the coupling of edge states in narrow ribbon, a band gap could be opened for specific spin and the time-reversal symmetry could be broken, leading to a spin polarized quantum anomalous Hall phase. Various kinds of spin-valley polarized edge states are formed at the two boundaries. Furthermore, the spin-valley polarized insulating states can be used to realize a perfect spin-valley switch.
\end{abstract}
\maketitle

\section{Introduction}

The discovery of graphene initiated an extensive search for other two-dimensional ($2$D) monolayer materials, and the monolayer topological materials in particular. As the close relatives of graphene, silicene \cite{Lalmi, Vogt, Fleurence}, germanene \cite{Gao, Davila}, and stanene \cite{Zhu} with atomically thin layer have been experimentally grown on different substrates in the past few years. Compared to graphene, silicene, germanene and stanene have a large intrinsic spin-orbit coupling (SOC) \cite{Yao, Zhang} and the band gap can be controlled by applying an electric field due to the buckled structure \cite{Drummond}. These $2$D honeycomb lattices can be described by the Hamiltonian of graphene with intrinsic SOC. Therefore, these materials are expected to be quantum spin Hall (QSH) insulators which characterized by an insulating bulk and topologically protected gapless edge states \cite{Hasan, Zhang2}. Although the QSH effect was first predicted in graphene \cite{Kane, Sun, Sun2}, it can occur only at unrealistically low temperatures due to the rather weak SOC \cite{Min, Yao2}. Recently, various detectable topological phases are proposed in silicene, germanene, and stanene, including the QSH effect \cite{Yao3, Yao4, Matusalem}, the quantum anomalous Hall (QAH) effect \cite{Ezawa, Hsu, PLi, Zou}, valley-polarized QAH effect \cite{Yao5, Zhou, Zhai}, and quantum spin-quantum anomalous Hall effect \cite{Ezawa2}. In particular, the inversion symmetry breaking could lead to a quantum valley Hall effect characterized by Berry phase effect and valley Chern number \cite{Xiao}. As an application, the edge state in the QSH system is important to the low-dissipation devices, and it can be used in spin filter and field-effect topological quantum transistor \cite{Ezawa3, Liu}.

With the advent of monolayer materials, spintronics and valleytronics have developed rapidly \cite{Schaibley, YLiu, Avsar}. Valleytronics exploits valley degree of freedom and its potential applications. Similar to spintronics, the central issue of valleytronics is the generation and manipulation of valley-polarized currents. A lot of researches on the control of spin and valley dependent quantum transport in QSH insulators have been reported \cite{Wang, Yao6, Tao, Rzeszotarski, Sun3, Ezawa4, Orellana, Lu, Sun4, Sun5, An, Jalil, Shakouri, ZYu, Zheng, Fiori, Lu2, Ezawa5, Tabert, Lee, Ominato}. In particular, the external exchange field and electric field are often employed on zigzag nanoribbons to control the spin and valley degrees of freedom. Recently, Tao et al. found a new way of generating spin-polarized current in QSH insulators by tuning the phase difference between spin up and down electrons, conceptually different from the conventional all-electrical approaches \cite{Wang}. Considering the extrinsic Rashba SOC, silicene can host topologically protected spin- and valley-polarized edge states which can be switched by reversing the electric field \cite{Yao6}. In zigzag germanene nanoribbon with thermal leads subjected to local noncollinear exchange fields, a thermoelectric spin-current generator was theoretically proposed \cite{Zheng}. Taking advantage of the modulation of the band gap and the edge-localized nature of the conduction- and valence-band states, a highly polarized spin currents could be achieved in stanene nanoribbons \cite{Fiori}. In addition, spintronics has been extended to spin-valleytronics by incorporating the valley degree of freedom \cite{Ezawa2, Ezawa5, Tabert, Lee}, such as spin-valley optical selection rule \cite{Ezawa5} and dissipationless spin-valley current \cite{Lee}. Tabert and Nicol demonstrated that four distinct valley- and spin-polarized levels and currents can be generated in silicene and other similar $2$D crystals by tuning the electric and magnetic fields \cite{Tabert}.

In previous literatures, the external fields are generally applied to the whole nanoribbon. Distinctively, in this work, we place the local potential and exchange fields to the boundaries of nanoribbon, such as silicene, germanene, and stanene. In fact, the study on graphene demonstrates that the side potential is an effective method of engineering the electronic structure \cite{Bhowmick, Apel, Chiu, Lee2}. When the side potential is antisymmetric, the energy spectrum of graphene nanoribbons could open up a gap \cite{Apel}. However, the study of side potential on the spin and valley dependent Hall effect in the nanoribbons with a spin-orbit coupling is still blank. The schematic diagram of the considered side potential is shown Fig. 1(a), which could effectively control the helical edge states along the zigzag nanoribbon. We demonstrate that the quantum spin-valley Hall states, valley polarized QSH states, spin polarized QAH states, and spin-valley polarized insulating states could appear by tuning the side potential and the ribbon width. In particular, the coupling of edge states could open a band gap and the side potential breaks the time-reversal symmetry. As a consequence, the QAH effect could be realized in the narrow ribbon. The system could achieve a remarkable spin and valley polarized current, where the current of certain spin is from the certain valley.

The paper is organized as follows. In Sec. II we introduce the considered tight-binding model and the nonequilibrium Green's function method for calculating energy band and conductance. The numerical results on the band structure and edge states controlled by side potential for the wide and narrow ribbons are presented in Sec. III. We conclude with a summary in Sec. IV.

\section{Theoretical Formulation}

\begin{figure}
\includegraphics[width=8.0cm,height=3.0cm]{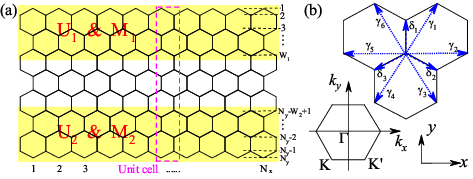}
\caption{ (a) Schematic plot of a zigzag nanoribbon system with potential field $U_{1,2}$ and exchange field $M_{1,2}$ along the boundaries. The ribbon width $N_y$ is defined by the number of silicon atoms in the unit cell. $W_{1,2}$ describes the width of the side potentials $U_{1,2}$ and $M_{1,2}$. The unit cell of the nanoribbon is marked by the dashed lines. (b) The nearest neighbor bonds and the next nearest neighbor bonds marked by $\delta_i$ and $\gamma_i$ in the lattice structure, respectively. The first Brillouin zone in the momentum space is also plotted. $K$ and $K'$ are the Dirac points.}
\end{figure}

Electrons in $2$D honeycomb lattices with an intrinsic SOC can be well described by the Kane-Mele tight-binding model \cite{Yao, Kane, Ezawa4}. Considering the side potentials $U_{1,2}$ and $M_{1,2}$, the Hamiltonian reads
\begin{eqnarray}
H = && -t \sum_{\langle i,j \rangle, \alpha} c_{i \alpha}^{\dag} c_{j \alpha} + i \frac{\lambda_{SO}}{3\sqrt{3}}
    \sum_{\langle\langle i,j \rangle\rangle, \alpha, \beta} v_{ij} c_{i \alpha}^{\dag} (\sigma_z)_{\alpha\beta} c_{j \beta} \nonumber\\
&&  + \sum_{i=1}^{W_{1}} \sum_{\alpha} \left[U_1 c_{i \alpha}^{\dag} c_{i \alpha}
+   M_{1} c_{i \alpha}^{\dag}
\left(\sigma_z\right)_{\alpha\alpha} c_{i \alpha}\right] \nonumber\\
&&  + \sum_{i=N_y-W_2+1}^{N_y} \sum_{\alpha} \left[ U_2 c_{i \alpha}^{\dag} c_{i \alpha}
+  M_{2} c_{i \alpha}^{\dag}
\left(\sigma_z\right)_{\alpha\alpha} c_{i \alpha}\right].
\end{eqnarray}
The first term describes the nearest-neighbor hopping with the hopping energy $t$, and $c_{i \alpha}^{\dag}$ ($c_{i \alpha}$) is the electronic creation (annihilation) operator with spin $\alpha$ ($\alpha=\uparrow,\downarrow$) at site $i$. The second term is intrinsic SOC with strength $\lambda_{SO}$, which involves spin dependent next nearest-neighbour hopping. $v_{ij} = +1 (-1)$ if the next nearest-neighboring hopping is anticlockwise (clockwise) with respect to the positive $z$ axis. $\sigma_z$ is the Pauli matrix associated with spin degree of freedom. As shown in Fig. 1(b), the nearest-neighbor bonds and the next nearest-neighbor bonds are marked by $\delta_i$ and $\gamma_i$ in the lattice structure, respectively. $\langle i, j \rangle$ and $\langle\langle i, j \rangle\rangle$ denote the sum over the nearest-neighbor and the next nearest-neighbor hopping sites, respectively. The third and last terms represent the potential field $U_{1,2}$ and exchange field $M_{1,2}$, respectively, which is applied to the lattice sites belonging to the boundaries, as shown in Fig. 1(a). The side potentials $U_{1,2}$ and $M_{1,2}$ are applied along two boundaries of zigzag nanoribbon with the width $W_{1,2}$, which can be induced by the gate voltages and the ferromagnetic insulators in experiment, respectively. Experimentally, a local exchange field with nanoscale on the 2D honeycomb lattices can be induced by the proximity effect, the value of which could reach to tens of $meV$ \cite{Jozsa, Yang}. Local gate control of the electrostatic potential with nanoscale and hundreds of $meV$ in nanoribbon-based devices has also been achieved by many experimental groups \cite{Williams, Huard, Ozyilmaz}. The Hamiltonian (1) is generally used to describe silicene, germanene, and stanene. It is predicted that the hopping energy $t=1.6eV$ and SOC $\lambda_{SO}=3.9meV=0.0024t$ for silicene, and $t=1.3eV$ and $\lambda_{SO}=43meV=0.033t$ for germanene \cite{Yao}. The first-principles calculations revealed that $t=1.3eV$ and $\lambda_{SO}=0.1eV=0.077t$ for stanene \cite{Zhang}. We take $\lambda_{SO}=0.05t$ and $t=1.3eV$ as an example in the following calculation. Changing the value of $\lambda_{SO}$ does not affect the qualitative results. Therefore, the discussion and conclusion are applicable to silicene, germanene, and stanene.

For the case of an infinite nanoribbon, the honeycomb structure along the $x$ direction is assumed to be periodic. The longitudinal wave vectors $k_x$ is a good quantum number which satisfies $[H, k_x ]=0$. Therefore, based on the tight-binding model and Bloch's theorem, the band structure of an infinite nanoribbon can be calculated, the $k_x$-dependent Hamiltonian of which can be written as
\begin{eqnarray}
H (k_x) = H_{00} + H_{01} e^{ik_xa} + H_{-10} e^{-ik_xa},
\end{eqnarray}
where $H_{00}$ is a unit cell Hamiltonian matrix of one chain, $H_{01}$ (or $H_{-10}$) is the coupling matrix with the right-hand (or left-hand) adjacent cell, and $a$ is the lattice constant.

The conductance $G$ for an electron with energy $E$ through the ribbon can be calculated by the nonequilibrium Green's function method. A two-terminal system comprises the central region, the left and right semiinfinite leads. The electrical conductance can be calculated by means of the Landauer-B\"{u}ttiker formula as \cite{Sun, Long}
\begin{eqnarray}
G(E) = \frac{e^2}{h} Tr [\Gamma_L(E) G^r(E) \Gamma_R(E) G^a(E)],
\end{eqnarray}
where $\Gamma_{L,R}(E) = i [\Sigma_{L,R}(E) - \Sigma^{\dag}_{L,R}(E)]$ is the linewidth function,
$G^r(E)$ and $G^a(E)$ are the retarded and advanced Green's functions given by $G^r(E) = [G^a(E)]^{\dag} = 1 / (E - H_c - \Sigma_L - \Sigma_R)$ with the Hamiltonian $H_c$ in the central region. $\Sigma_{L,R}$ is the selfenergy caused by the coupling between the central and lead regions. The selfenergy function can be calculated by $\Sigma_L = H_{CL} g_L H_{LC} = H_{CL} g_L H_{CL}^{\dag}$ and $\Sigma_R = H_{CR} g_R H_{RC} = H_{CR} g_R H_{CR}^{\dag}$,  where $H_{CL}$ and $H_{CR}$ are the Hamiltonian matrices of the coupling between the central region and lead regions. $g_{L,R}$ is the surface Green's function which can be calculated using iteration method \cite{Datta}.

\section{Results and Discussions}

Is this section, we mainly study the impact of side potential with different symmetries on the band structure, edge states, and transport property. In addition, the influence of finite size effect on the edge states may be significant for QSH insulator and QAH insulator\cite{Zhou2, Fu}, which can produce an energy gap in the HgTe/CdTe quantum well. In narrow zigzag graphene nanoribbons, a braiding of the conduction and valence bands can be generated by the third-neighbor hopping \cite{Correa}. Thus, we also study the size effect of nanoribbon on the band structure and edge states, which are crucial for the device application. We consider a wide ribbon in Sec. III A and a narrow ribbon in Sec. III B and III C.

In order to distinguish the edge states at the upper and lower boundaries for different spin electrons, an antisymmetric potential field is considered, which is fixed as $U_1=-U_2=U$, while the exchange field may be symmetric or antisymmetric. Note that the widths $W_{1,2}$ of the upper and lower potentials are assumed to be the same for convenience in the following discussion, i.e., $W_1=W_2=W$. When the widths of the upper and lower potentials become different, i.e., $W_1 \neq W_2$, the energy band with respect to $E=0$ will become asymmetric slightly, however, the main conclusion on various Hall effects is still valid. In addition, for purpose of clearly seeing the band for edge states, we only present the local profile of the band in the figures. The $K$ and $K'$ points are located at $k_x=2\pi/3a$ and $k_x=4\pi/3a$, respectively, while the $\Gamma$ point is located at $k_x=0$ [see Fig. 1(b)]. In the following figures, the band structures (or the edge states) for spin-up and spin-down electrons are highlighted by the red and blue curves (or arrows), respectively, which can be calculated by Eq. (2). The hopping energy $t$ is chosen as the energy unit in the calculation.

\subsection{Spin-valley polarized edge states in wide ribbon}

\begin{figure}
\includegraphics[width=8.0cm,height=7.0cm]{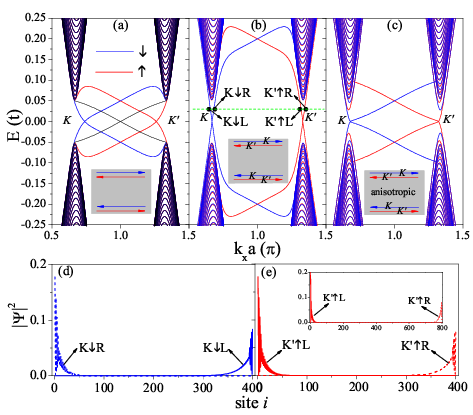}
\caption{ Band structures of the wide zigzag ribbon with (a) $W=8$ and $U=0.05t$; (b) $W=8$ and $U=0.2t$; (c) $W=80$ and $U=0.05t$. The black thin curve in (a) is the band structure of a pristine ribbon, i.e., $U=0.0$. The red and blue curves are for spin-up and spin-down electrons, respectively. The red and blue arrows denote the propagation directions of the opposite spins at the boundaries. (d)(e) Probability density of wave function $|\Psi|^2$ for the edge states labeled by black points in (b) as a function of the site $i$ in the $y$ direction. Other parameters are set as $N_y=400$ and $M_1=M_2=0.0$. The inset in (e) is $|\Psi|^2$ for the ribbon width $N_y=800$ and other parameters being the same as that in (e).}
\end{figure}

Firstly, we discuss the effect of side potential in the wide nanoribbon and take the width $N_y=400$ as an example. The result is more significant for wider nanoribbons. Fig. 2 shows (a)-(c) the band structure and (d)(e) the edge states when only the potential field is applied, i.e., $M_1 = M_2 = 0.0$. The black thin curve in Fig. 2(a) represents the energy band for pristine ribbon with $U=0.0$ which is spin degenerate, characterized by the gapless edge states. The spin-up state at lower boundary and the spin-down state at upper boundary are degenerate with a forward move, while the spin-up state at upper boundary and the spin-down state at lower boundary are degenerate with a backward move. However, with the appearance of $U$, the bands for spin-up and spin-down states at upper boundary are shifted up along the energy, while the bands for spin-up and spin-down states at lower boundary are shifted down. Consequently, the spin degeneracy is lifted due to the combined effect of the potential field and the intrinsic SOC, as shown in Fig. 2(a). As $U$ increases, the band crossing point splits, and the crossing point for spin up move to right, oppositely, the crossing point for spin down move to left [see Fig. 2(a)]. With the further increase of $U$, the band for spin-up (or spin-down) edge states would move to the $K'$ (or $K$) valley [see Fig. 2(b)]. Thus, the system is not only QSH insulator but also quantum valley Hall insulator, that is quantum spin-valley Hall effect. Figs. 2(d) and 2(e) display the probability density of wave function for the spin-down and spin-up edge states, respectively, which are labeled by black points in Fig. 2(b). One can clearly see that the spin-valley polarized edge states are formed at the two boundaries, where the edge states are spin- and valley-polarized simultaneously. Thus, the currents carried by the upper and lower edge states are spin-valley polarized, which could be extracted by a local terminal. Note that for a wider ribbon, the localization of the edge states will be more remarkable [see the inset in Fig. 2(e)]. Obviously, the side potential mainly affects the band of edge state and has very little effect on the band of the bulk state. The reversal of $U$ can interchange the role of spin-up and spin-down electrons. In addition, the comparison between Figs. 2(a) and 2(c) indicates that the width of $U$ can be used to control the dispersion in the wave vector space and to modulate the Fermi velocity of spin-up and spin-down electrons. It can be seen that the Fermi velocities $v= \partial E / \partial k$ near the Fermi level $E=0$ of forward (or backward) move for spin up and spin down at the two valleys are different and the edge modes are anisotropic helical. It is suggested that the time reaching the steady state are different for spin up and spin down. Therefore, the conductance contributed by the anisotropic edge states can be spin and valley polarized before reaching the steady state \cite{Wang}.

\begin{figure}
\includegraphics[width=8.0cm,height=4.0cm]{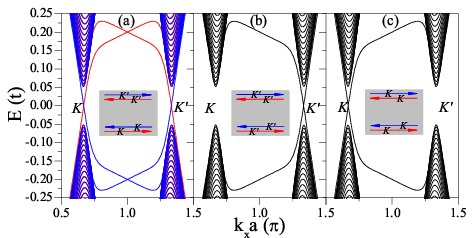}
\caption{ Band structures of the wide zigzag ribbon with (a) $M_1=M_2=M=0.2t$, (b) $M_1=-M_2=M=0.2t$, and (c) $M_1=-M_2=M=-0.2t$. Here, $N_y=400$, $U=0.0$, and $W=8$. }
\end{figure}

Fig. 3 discusses the band structure when only exchange field is applied with $U=0.0$, which is distinct from the one observed in Fig. 2. For a symmetric exchange field with $M_1=M_2=M$, the bands for spin-up states at both boundaries shift up, inversely, the bands for spin-down states at both boundaries shift down. As shown in Fig. 3(a) at $M=0.2t$, the spin splits and the crossing points move to the two valleys. The system always hold the QSH insulator with valley-polarized edge states, leading to the valley-polarized QSH effect. More interestingly, the electrons at the $K$ valley just motion on the lower boundary, while the electrons from the $K'$ valley motion on the upper boundary, and so the system can work as a spin-valley filter. Fig. 3(b) exhibits the band structure under an antisymmetric exchange field with $M_1 = -M_2 = M$. In this case, for the four helical edge states, the band of states moving to the left (or right) would shift up (or down), and so the spin remains degenerate. With the increase of $M$, the crossing point moves to the right. For a proper value of $M$ such as $M=0.2t$, the valley-polarized edge states are obtained at $K'$ valley and a energy gap is resulted at $K$ valley. As a result, another kind of valley-polarized QSH effect is obtained. Note that a valley-polarized edge state at $K$ valley can also be formed when $M=-0.2t$ [see Fig. 3(c)]. Therefore, the conductance in the low energy region is completely contributed by the edge states at the $K$ (or $K'$) valley, and one can achieve a perfect valley-polarized conductance by exchange field.

\subsection{Quantum anomalous Hall states in narrow ribbon}

\begin{figure}
\includegraphics[width=8.0cm,height=4.0cm]{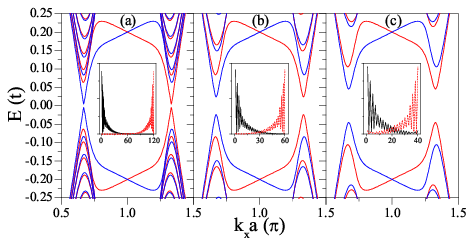}
\caption{ Band structures of the narrow zigzag ribbon with (a) $N_y=120$, (b) $N_y=60$, and (c) $N_y=40$. Here, $U=0.2t$, $M_1=M_2=0.0$, and $W=8$. The insets show the probability density of wave function for spin up at $k_x a=1.3\pi$ near Dirac cone. }
\end{figure}

Next, we turn to discuss the nanoribbon with a narrow width $N_y$. Fig. 4 presents the band structure with $W=8$ for different values of $N_y$. One may find that an observable band gap is opened up at the Dirac points. When the ribbon width becomes narrow, the gap is broadened gradually. Such a phenomenon arises from the effect of side potential as well as the coupling of the edge states at the upper and lower boundaries via tunnel effect. The insets of Fig. 4 show the probability density of wave function for spin up at $k_x a = 1.3\pi$ corresponding to the lowest minibands near the Fermi level. We can see that as $N_y$ decreases, the coupling of the edge states becomes stronger and stronger, and so the gap is enlarged.

\begin{figure}
\includegraphics[width=8.0cm,height=6.0cm]{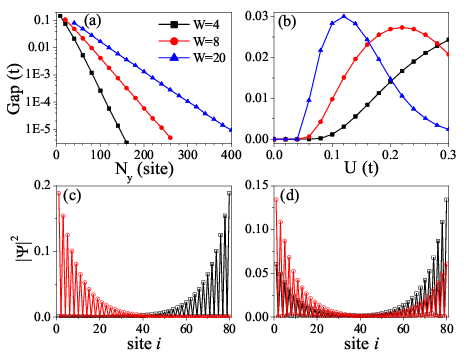}
\caption{ Band gap versus (a) ribbon width $N_y$ with $U=0.1t$ and (b) potential field $U$ with $N_y=80$. Probability density of wave function for spin up with (c) $U=0.04t$ and (d) $U=0.08t$ at $N_y=80$, $W=8$, and $k_x a=1.3\pi$. Here, $M_1=M_2=0.0$. }
\end{figure}

The dependence of band gap on (a) the ribbon width $N_y$ and (b) the potential field $U$ is shown in Fig.5 with different values of width $W$. As ribbon width $N_y$ increases, the band gap is decreased because the coupling of the edge states at the two boundaries becomes weak [see Fig. 5(a)]. On the other hand, from Fig. 5(b) one can clearly see that the gap is zero and the system keeps the QSH insulator at $U<\lambda_{SO}$. As $U$ further increases, a spectral gap can be induced, the edge states disappear, and the system is driven to trivial band insulator (BI). Subsequently, the gap could be closed again at a certain value of $U$, and then the system converts metal. Thus, the system undergoes a phase transition from QSH insulator phase to BI phase and then to metal phase controlled by $U$, which is distinct from graphene. $U=\lambda_{SO}$ is the critical value between QSH insulator phase and BI phase.

In fact, the electron motion in the low energy region can be described by a two-band Hamiltonian,
\begin{eqnarray}
H = \hbar v_F ( k_x \tau_x - \eta k_y \tau_y )  - \eta \sigma \lambda_{SO}  \tau_z + U - \sigma M_{1,2},
\end{eqnarray}
where $\eta = \pm 1$ denotes the $K$ and $K'$ valleys and $\sigma = \pm 1$ denotes spin-up and spin-down states. The eigenvalue of Eq. (4) can be written as $ E = \pm  \sqrt{ \lambda^2_{SO}  + ( \hbar v_F)^2 ( k_x + k_y )^2} +U - \sigma M_{1,2} $. Then the transverse wave vector reads $ k_y = \sqrt{ ( E - U + \sigma M_{1,2} )^2 -  \lambda^2_{SO} - ( \hbar v_F k_x )^2 } / \hbar v_F $ which determines the nature of electronic states. Because $k_x$ is the good quantum number and conserved quantity, the wave functions in the different regions have the form: $\psi=A e^{i k_y y} + B e^{-i k_y y} $ \cite{Apel}. General properties of the solution for $i k_y$ determines the distribution of the wave functions in ribbon, which can be adjusted by $U$ and $M_{1,2}$. When $U<\lambda_{SO}$ with $M_1=M_2=0.0$, the wave vector $k_y$ is invariably imaginary and the eigenstate is evanescent wave near the Fermi level $E=0.0$. Consequently, the edge states at one boundary cannot tunnel to another boundary and couple with other edge states, and so the band gap cannot be opened up. For $U>\lambda_{SO}$, $k_y$ is real and the eigenstate is traveling wave at the two boundary. Then the edge states could tunnel and couple with other edge states, leading  to the band gap. As shown in Figs. 5(c) and 5(d), the probability density of wave function for spin-up edge states is plotted at $k_x a=1.3\pi$. When $U=0.04t<\lambda_{SO}$ in Fig. 5(c), the edge state at the upper boundary is localized at the $a$ sublattice (odd number site), while the edge state at the lower boundary is localized at the $b$ sublattice (even number site), and there is exactly no overlap between the two edge modes. Thus, the coupling between them is very weak. When $U=0.08t>\lambda_{SO}$ in Fig. 5(d), the edge state at $a$ (or $b$) sublattice of the upper (or lower) boundary tunnels to the $b$ (or $a$) sublattice of the lower (or upper) boundary and couple with each other, generating a gap in the band. Based on the the boundary conditions and the energy dispersion, it is found that a finite energy gap opens which is approximately proportional to $e^{- k_y N_y}$ \cite{Zhou2}. It is suggested that the gap decays in an exponential law of $N_y$, i.e., the logarithm for the gap decays linearly with $N_y$, as shown in Fig. 5(a). The gap could reach $0.1t$ for the narrow ribbon. However, for a wider ribbon, the coupling of the edge states is very weak even though $U>\lambda_{SO}$ and the gap scale is less than $10^{-7}t$ trending to zero [see Figs. 5(a) and 2(b)]. For the antisymmetric exchange field, the same results on band gap can be obtained as shown in Figs. 5(a) and 5(b). In fact, for the spin-up (or spin-down) electrons, the potential energy under antisymmetric exchange field is completely the same (or opposite) as the one under antisymmetric potential field. This results in an identical energy gap in both fields.

\begin{figure}
\includegraphics[width=8.0cm,height=4.0cm]{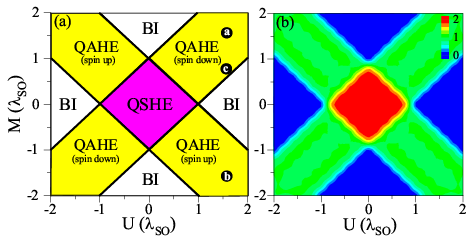}
\caption{ (a) Phase diagram of the narrow zigzag ribbon in the $(U,M)$ space. (b) Contour map of conductance $G(U, M)$ at $N_y=48$, $W=12$, and $E=0.0$. The length of the central region is $N_x=100$. $e^2/h$ is the conductance unit of $G$; $\lambda_{SO}$ is the energy unit of $U$ and $M$ in (b). Here, $M_1=-M_2=M$.}
\end{figure}

Discussions in Sec. III A demonstrate that the effect of antisymmetric exchange field on the band structure of edge states is different from that of antisymmetric potential field [see Figs. 2(b) and 3(b)] for the wide ribbon. However, in the narrow ribbon, both antisymmetric potential field and antisymmetric exchange field could induce a band gap. Although the size of band gap is spin-independent, the band structure is spin-dependent. Taking advantage of a joint control of the potential and exchange fields, the coupling between the edge states will become spin dependent. Therefore, one can open a gap for a specified spin index and destroy its edge states while another spin keeps gapless, where the time reversal symmetry is broken and the spin degeneracy is lifted. Fig. 6(a) presents a phase diagram of the narrow zigzag ribbon in the $(U,M)$ space when the side potential composed of antisymmetric potential and exchange fields, where $U_1=-U_2=U$ and $M_1=-M_2=M$. The effective side potentials on the upper and lower boundaries are $U+\sigma M$ and $-U-\sigma M$, respectively. When $|U+\sigma M|<\lambda_{SO}$, the system has no gap for both spins and exhibits QSH effect, corresponding to the magenta region in Fig. 6(a). When $|U+M|<\lambda_{SO}$ but $|U-M|>\lambda_{SO}$ (or $|U-M|<\lambda_{SO}$ but $|U+M|>\lambda_{SO}$), the spin-down (or spin-up) electron would open a band gap and lose the edge states, while the edge states for spin-up (or spin-down) electron always exist at the boundaries. As a consequence, a spin polarized QAH effect is realized, corresponding the yellow region in Fig. 6(a). When $|U+\sigma M|>\lambda_{SO}$, both spins have a gap and the system becomes BI, labeled by the white region in Fig. 6(a).

\begin{figure}
\includegraphics[width=8.0cm,height=6.0cm]{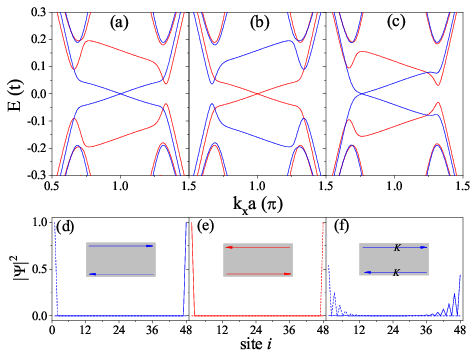}
\caption{ Band structures of the narrow zigzag ribbon labeled by black points in Fig. 6(a). (a) $U=0.08t$ and $M_1=-M_2=M=0.08t$; (b) $U=0.08t$ and $M_1=-M_2=M=-0.08t$; (c) $U=0.08t$ and $M_1=-M_2=M=0.04t$. (d)-(f) show the probability density of wave function $|\Psi|^2$ for the edge states at $E=0.001t$ in (a)-(c), respectively. Here, $N_y=48$ and $W=12$. }
\end{figure}

Figs. 7(a)-7(c) display the band structures of the narrow ribbon with $N_y=48$ labeled by black points in the phase diagram of Fig. 6(a). For a special case, i.e., $U=M$, the effective potential for spin down is zero, while the potential for spin up is $\pm U \pm M$. As expected, there are only two spin-down edge states and spin-up edge states are destroyed, confirming that the system becomes a spin-down polarized QAH insulator [see Fig. 7(a)]. The spin-up polarized QAH states can also be achieved by changing the sign of side potential [see Fig. 7(b)]. Fig. 7(c) presents the band structure at the boundary between the QAH effect and BI. We can find that the crossing point of spin down move to the $K$ valley and the spin up has a band gap, suggesting that the QAH effect is not only spin polarized but also valley polarized. Figs. 7(d)-7(f) show the probability density of wave function $|\Psi|^2$ for the edge states at $E=0.001t$ in (a)-(c), respectively. There are only two edge states from a specific spin which is localized at $a$ (or $b$) sublattice of the upper (or lower) boundary.

By connecting the system with two leads consisting of zigzag nanoribbons, we can study the transport behaviors in QAH insulator using Eq. (3). The $x$ direction is finite and its length is $N_x=100$ in the calculation for the transport. The contour map of conductance $G(U,M)$ is shown in Fig. 6(b) when ribbon width is $N_y=48$ and Fermi energy is $E=0.001t$. Remarkably, the conductance in Fig. 6(b) is completely consistent with the phase diagram in Fig. 6(a). In the QSH effect region, the conductance is $2 e^2/h$ contributed by the helical edge states. In the QAH effect region, the conductance is $e^2/h$ contributed by the chiral edge states of one spin and the conductance for another spin is zero, resulting in a remarkable spin polarized conductance.

\subsection{Spin-valley switch in narrow ribbon}

\begin{figure}
\includegraphics[width=8.0cm,height=4.0cm]{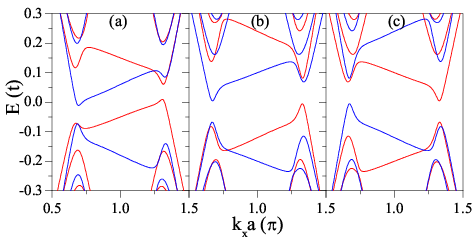}
\caption{ Band structures of the narrow zigzag ribbon with (a) $U=0.1t$ and $M_1=M_2=M=0.05t$; (b) $U=0.15t$ and $M_1=M_2=M=0.05t$; (c) $U=0.15t$ and $M_1=M_2=M=-0.05t$. Here, $N_y=48$ and $W=12$. }
\end{figure}

The potential field or the exchange field alone cannot generate a spin polarized current in two-terminal device, since the band structures for the two spins are either antisymmetric [see Figs. 2(a)-2(c)], or symmetric [see Fig. 3(a)], or degenerate [see Figs. 3(b) and 3(c)]. However, the combined effect of the potential and exchange fields could induce a spin and valley polarization in transport. In Fig. 8, the band structure under the side potential composed of antisymmetric potential field and symmetric exchange field is discussed. In this case, the effective potentials at upper and lower boundaries are $U+\sigma M$ and $-U+\sigma M$, respectively. The potential difference between the upper and lower boundaries is $2U$ for both spins. When $U>\lambda_{SO}$, the band gaps for both spin can be opened up simultaneously. $M$ determines the symmetry of the band structure [see Figs. 8(b) and 8(c)]. As a result, the band structure for spin-up and spin-down states becomes asymmetric and the band gap is generated. Figs. 8(a) and 8(b) indicate that the gap can be enlarged properly by adjusting $U$. In the vicinity of Fermi level, the band for spin up (or spin down) is concentrated at $K'$ (or $K$) valley. Consequently, the spin and valley polarized insulating states are realized near Fermi level, leading to the spin and valley polarized conductance in Fig. 9.

\begin{figure}
\includegraphics[width=8.0cm,height=6.0cm]{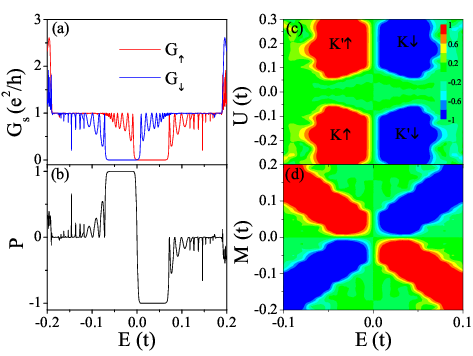}
\caption{ (a) Spin dependent conductance and (b) spin polarization versus Fermi energy $E$, and the parameter values are the same as these in Fig. 8(b). (c)(d) Contour map of spin polarizations (c) $P(E,U)$ with $M_1=M_2=M=0.05t$ and (d) $P(E,M)$ with $U=0.15t$ and $M_1=M_2=M$. The length of the central region is $N_x=100$. }
\end{figure}

Fig. 9 shows (a) the spin dependent conductance and (b) its spin polarization $P=(G_{\uparrow}-G_{\downarrow})/(G_{\uparrow}+G_{\downarrow})$, corresponding to the band structure in Fig. 8(b). By comparing Figs. 8(b) and 9(a), one may find that the conductance $G_{\uparrow}$ is mainly offered spin-up electron at $K'$ valley and $G_{\downarrow}$ is offered spin-down electron at $K$ valley. Thus, a spin and valley polarized conductance is realized where both spin and valley degrees of freedom are polarized simultaneously [see Fig. 9(b)]. By adjusting the potential field $U$ and Fermi energy $E$, one could achieve a polarized conductance contributed by certain spin from certain valley. The oscillations of $G_{\uparrow, \downarrow}$ and $P$ result from the resonant mode in the scattering region where the side potentials are applied. Figs. 9(c) and 9(d) show the contour plot of spin polarizations $P(E, U)$ and $P(E, M)$, respectively. It can be seen that there is no polarization in the absence of $U$ or $M$. When $U>\lambda_{SO}$, perfect polarization platform for certain spin from certain valley could be realized by the side potential. Furthermore, the energy region for polarization platform could be controlled by applying exchange field [see Fig. 9(d)]. The result suggests that the system can work as a spin-valley switch.

\section{Conclusion}

In conclusion, we studied the band structure and edge states of the $2$D honeycomb lattices with intrinsic SOC controlled by the side potential and the finite size effect of nanoribbon. The side potential is made up of potential field and exchange field. For a wide ribbon, the quantum spin-valley Hall phase by the antisymmetric potential field and the valley polarized QSH phase by the symmetric or antisymmetric exchange field could be proposed, leading to the spin-valley polarized edge states. For the narrow ribbon, the coupling of edge states would become very strong, and so the side potential composed of antisymmetric potential field and antisymmetric exchange field could produce a band gap for a certain spin. As a result, the system becomes spin polarized QAH insulator. In addition, an effective spin-valley switch could be realized by antisymmetric potential field and symmetric exchange field. These results should be conducive to the potential applications of the spin and valley polarized edge states.

This work was supported by the NSFC (Grants No. 11974153 and No. 11921005), National Key R and D Program of China (Grant No. 2017YFA0303301), and the Strategic Priority Research Program of Chinese Academy of Sciences (Grant No. XDB28000000). Email address of Wei-Tao Lu: physlu@163.com.

%\end{CJK*}
\end{document}